\documentclass[a4paper,pre,twocolumn,amsmath,amssymb,superscriptaddress,longbibliography]{revtex4-1}
\usepackage{graphicx,color}
\usepackage{soul}
\usepackage[colorlinks,linkcolor=magenta,citecolor=magenta]{hyperref}

\newcommand{\dd}{\mathrm{d}}
\newcommand{\td}[2]{\frac{\dd #1}{\dd #2}}
\newcommand{\pd}[2]{\frac{\partial #1}{\partial #2}}

\newcommand{\unit}[1]{\;\mathrm{#1}}

\newcommand{\al}{\alpha}

\newcommand{\eps}{\varepsilon}
\newcommand{\kap}{\kappa}

\newcommand{\eb}{\eps_\text{b}}
\newcommand{\kT}{k_\text{B}T}

\AtBeginDocument{%
    \newwrite\bibnotes
    \def\bibnotesext{Notes.bib}
    \immediate\openout\bibnotes=\jobname\bibnotesext
    \immediate\write\bibnotes{@CONTROL{REVTEX41Control}}
    \immediate\write\bibnotes{@CONTROL{%
    apsrev41Control,author="08",editor="1",pages="1",title="0",year="1"}}
     \if@filesw
     \immediate\write\@auxout{\string\citation{apsrev41Control}}%
    \fi
}%

\begin{document}

\title{Inverse condensation of adsorbed molecules with two conformations}

\author{Jo\"el A. K. L. Picard}
\affiliation{Institut f\"ur Physik, Johannes Gutenberg-Universit\"at Mainz, Staudingerweg 7-9, 55128 Mainz, Germany}
\author{Thomas Speck}
\affiliation{Institute of Theoretical Physics IV, University of Stuttgart, 70569 Stuttgart, Germany}

\begin{abstract}
  Conventional gas-liquid phase transitions feature a coexistence line that has a monotonic and positive slope in line with our intuition that cooling always leads to condensation. Here we study the inverse phenomenon, condensation of adsorbed organic molecules into dense domains upon heating. Our considerations are motivated by recent experiments [Aeschlimann \emph{et al.}, Angew. Chem. (2021)], which demonstrate the partial dissolution of an ordered molecular monolayer and the mobilization of molecules upon cooling. We introduce a simple lattice model in which each site can have three states corresponding to unoccupied and two discernible molecular conformations. We investigate this model through Monte Carlo simulations, mean-field theory, and exact results based on the analytical solution of the Ising model in two dimensions. Our results should be broadly applicable to molecules with distinct conformations that have sufficiently different entropies or heat capacities.
\end{abstract}

\maketitle

%% ---- introduction ----

\section{Introduction}

How a collection of disordered molecular building blocks autonomously arranges into structured and functional materials is of broad interest, from the synthesis of novel materials to our understanding of living matter~\cite{whitesides02}. Statistical physics provides a powerful framework through which the emergence of structure is connected to the underlying physics of phase transitions. Examples range from the ordering of lipids into membranes and tubules~\cite{thomas95} to the drying-induced self-assembly of nanoparticles on a surface~\cite{ge00}, the morphologies of which can be captured in a minimal lattice model~\cite{rabani03}. Morphologies of thin films of organic molecules deposited on a substrate~\cite{hlawacek08,kuhnle09,empting21} play an important role for organic electronics and photovoltaics~\cite{kippelen09}, in particular the control over complex arrested morphologies due to the intricate interplay of thermodynamic and kinetic factors~\cite{whitelam09,hagan11}.

Here we investigate a generic lattice model of adsorption sites, each of which can be occupied by a molecule. Our study is motivated by recent experimental studies of the structure formation of dimolybdenum tetraacetate, Mo$_2$(O$_2$CCH$_3$)$_4$, on a copper Cu(111) substrate in vacuum~\cite{kollamana18,aeschlimann21}. At room temperature, molecules at sub-monolayer coverage arrange into aligned chains. Remarkably, upon cooling the substrate the ordered domains partially dissolve and molecules become mobile again, which is reminiscent of inverse melting~\cite{greer00}. The idea of inverse freezing and melting has a long history starting with Tammann~\cite{tammann03} in 1903 but has been discussed mostly in connection with rather exotic systems such as helium above 20 bar~\cite{tedrow69}. Only rather recently it has been realized that adsorbed organic molecules are good candidates to observe inverse transitions more widely~\cite{scholl10}. These systems have in common that the liquid-solid phase boundary has an inflection point at high pressure and finite temperature so that further reducing the temperature leads to a reentrance into the liquid phase. Reentrant phase behavior, \emph{inter alia}, is also found for complex network fluids~\cite{tlusty00,russo11}, liquid crystals~\cite{cladis88}, and, generically, liquid mixtures~\cite{narayanan94,plazanet04}. In contrast, here we consider the phase behavior of a single molecular component.

\begin{figure}[t]
  \centering
  \includegraphics{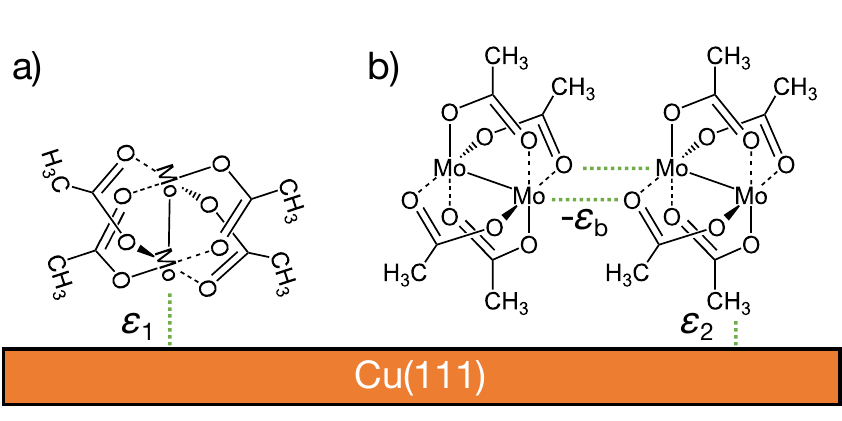}
  \caption{Chemical structure of dimolybdenum tetraacetate and the two tentative conformations on top of a copper Cu(111) substrate: (a)~Lying molecules binding with $\eps_1$. (b)~Standing molecules binding with $\eps_2$. In addition, intermolecular Mo--O bonding is assumed to contribute a lateral binding energy $-\eb$, which is absent for lying molecules. Not shown here but important for the model is that both conformations are supposed to have different entropies.}
  \label{fig:conf}
\end{figure}

At thermodynamic equilibrium, coexisting phases in an inhomogeneous system have to have the same chemical potential. Changing the intensive thermodynamic variables, along the coexistence line the change of chemical potentials thus has to be equal, $\dd\mu_\text{g}=\dd\mu_\text{l}$, using gas (g) and liquid (l) as a specific example. From the Gibbs-Duhem relation $\dd\mu_x=-s_x\dd T+v_x\dd p$ one immediately obtains the Clausius-Clapeyron relation
\begin{equation}
  \td{p}{T} = \frac{s_\text{g}-s_\text{l}}{v_\text{g}-v_\text{l}}
\end{equation}
for the slope of the coexistence line, where $s_x$ is the entropy and $v_x$ is the volume per molecule in the corresponding bulk phase, and $T$ and $p$ are conjugate temperature and pressure, respectively. Since we typically expect more available volume in the disordered gas phase together with a larger translational entropy, the slope is positive. A well-known exception is the solid-liquid transition of water exhibiting a negative slope since the volume of ice is larger. Another possibility is that the entropy of the more ordered phase is larger. Clearly, in this case the loss of translational entropy has to be offset by a gain of, e.g., intramolecular entropy. The minimal scenario thus involves two conformations of a complex organic molecule so that in the ordered (solid/liquid) phase the conformation with the larger entropy is found. For dimolybdenum tetraacetate, scanning tunneling microscopy reveals that ordered molecules are standing while freely diffusing molecules lie flat on the substrate (Fig.~\ref{fig:conf}), thus restricting intramolecular vibrations.

Schupper and Shnerb have introduced a spin model that highlights these minimal ingredients necessary for an inverse order-disorder transition~\cite{schupper04,schupper05}. It is based on the Blume-Capel model~\cite{blume66,capel66} with three states for each spin (conformation 1, unoccupied, conformation 2). This model features a tricritical point separating a line of critical points from a first-order transition line. Introducing a degeneracy that endows occupied sites with a higher internal entropy, the line of critical points develops an inflecting point and there is a reentrant behavior back into the disordered paramagnetic phase as temperature is reduced.

Informed by the experiments~\cite{aeschlimann21}, here we study an even simpler model that qualitatively reproduces the behavior observed in the experiments. As before we consider three states corresponding to unoccupied sites and the two conformations, but now only conformation 2 interacts laterally (Fig.~\ref{fig:conf}). This results in the Ising model for the interacting conformations, which condense into ordered domains below the critical temperature. The non-interacting lying molecules form a gas of mobile molecules outside these domains. Since molecules can freely convert between both conformations, only the total number of molecules is conserved (the so-called semi-grand canonical ensemble).

%% ---- model ----

\section{Model and methods}

\subsection{Lattice gas model}
\label{sec:lg}

We consider $N$ molecules adsorbed onto a two-dimensional substrate, the atomistic structure of which defines a regular lattice. The substrate is held at constant temperature $T$. We assume that each molecule can be found in one of two conformations with numbers $N_1$ and $N_2$ obeying $N_1+N_2=N$. Specifically for dimolybdenum tetraacetate, conformation 1 is lying on the substrate while conformation 2 is standing upright, cf. Fig~\ref{fig:conf}. Holding a single molecule fixed at a lattice site and in a given conformation $\al=1,2$ gives rise to the constrained partition functions
\begin{equation}
  \sum_{\xi\in\al} e^{-H(\xi)/\kT} = e^{-f_\al/\kT}
\end{equation}
defining the free energies $f_\al(T)$. Here, the sum is over all microstates $\xi$ compatible with the conformation $\al$, $H(\xi)$ is the Hamiltonian assigning each microstate an energy, and $k_\text{B}$ denotes Boltzmann's constant.

A coarse-grained configuration of the system is now determined by the positions of the $N$ molecules and their conformations. We assume that molecules bind laterally with binding energy $-\eb$ when both are in conformation 2 and neglect lateral binding otherwise. We assign each lattice site a variable $\hat n_i=1$ if site $i$ is occupied by a molecule in conformation 2 (standing) and $\hat n_i=0$ if the lattice site is unoccupied \emph{or} occupied by a lying molecule (which do not interact laterally). We thus arrive at the free energy functional
\begin{equation}
  \mathcal F(\{\hat n_i\};T,N) = -\eb\sum_{(ij)}\hat n_i\hat n_j + h(T)\sum_i\hat n_i + Nf_1
  \label{eq:F:lg}
\end{equation}
with $h(T)\equiv f_2-f_1$ the free energy difference of a single molecule in conformation 2 compared to conformation 1, which plays the role of a temperature-dependent external field. We recognize Eq.~\eqref{eq:F:lg} as the lattice gas (equivalent to the Ising model~\cite{lee52}) with a fluctuating number $N_2$ of molecules coupled to a finite ``reservoir'' with $N_1+N_2=N$ held fixed.

\subsection{Mean-field theory}

The mean-field free energy $F$ derived from Eq.~\eqref{eq:F:lg} is composed of two terms, the free energy of two ideal gases (corresponding to the two conformations) and an interaction term. We employ as order parameters
\begin{equation}
  x \equiv \frac{N_2}{N}, \qquad \phi \equiv \frac{N}{K}
\end{equation}
representing the fraction $0\leqslant x\leqslant 1$ of standing molecules and the coverage $0\leqslant\phi\leqslant 1$, where $K$ is the number of lattice sites (assuming for simplicity that both conformations occupy the same area). The free energy for $N$ non-interacting molecules
\begin{equation}
  F_0(N,K) = -TS_0 = \kT Kf_0(N/K)
\end{equation}
is given by the entropy 
\begin{equation}
  S_0(N,K) = k_\text{B}\ln{K\choose N} = k_\text{B}\ln\frac{K!}{N!(K-N)!}
\end{equation}
to distribute the molecules among $K$ lattice sites with
\begin{equation*}
  f_0(x) \simeq x\ln x + (1-x)\ln(1-x)
\end{equation*}
using Stirling's approximation. For the standing molecules there are $K$ sites with coverage $\phi_2\equiv N_2/K=x\phi$. The $N_1$ lying molecules can access only the remaining $K_1=K-N_2$ lattice sites. Adding both contributions and after some algebra we obtain
\begin{equation}
  F_0(N_1,K_1) + F_0(N_2,K) = \kT K\left[\phi f_0(x) + f_0(\phi)\right].
  \label{eq:F0}
\end{equation}
The interaction term in Eq.~\eqref{eq:F:lg} is approximated through
\begin{equation}
  -\eb\sum_{(ij)} \hat n_i\hat n_j = -\frac{1}{2}\eb\sum_i\hat n_i \sum_{j\in N_i}\hat n_j = -\frac{1}{2}\eb N_2\bar n_2,
  \label{eq:mf}
\end{equation}
where we have replaced the inner sum by the average number $\bar n_2=zN_2/K$ of standing neighbors with coordination number $z$ ($z=4$ for the square lattice). Adding this term to the ideal gas contribution Eq.~\eqref{eq:F0}, the reduced free energy density now reads
\begin{multline}
  f(x,\phi) \equiv \frac{F/K}{z\eb} \\
  = -\frac{1}{2}(x\phi)^2 + hx\phi + T\phi f_0(x) + Tf_0(\phi) + \phi f_1.
\end{multline}
Here and in what follows we employ dimensionless quantities through rescaling free energies $f_\al/(z\eb)\to f_\al$ and temperature $\kT/(z\eb)\to T$.

Since molecules can interconvert between both conformations, the first equilibrium condition is
\begin{equation}
  \pd{f}{x} = \left(-\phi x + h + T\ln\frac{x}{1-x}\right)\phi = 0,
\end{equation}
which yields the mean-field equation of state
\begin{equation}
  \phi(x;T) = \frac{1}{x}\left[h + T\ln\frac{x}{1-x}\right].
  \label{eq:phi}
\end{equation}
Above the critical temperature, $T>T_\text{c}$, the curve $\phi(x;T)$ is monotonic. For a fixed $\phi$ there is one composition $x$ and the system remains homogeneous.

Below the critical temperature, the system becomes inhomogeneous. At coexistence, both the reduced chemical potential
\begin{equation}
  \mu = \frac{1}{z\eb}\pd{F}{N} = \pd{f}{\phi}
\end{equation}
and the reduced pressure 
\begin{equation}
  p = -\frac{a}{z\eb}\pd{F}{A} = -(f-\phi\mu)
\end{equation}
need to be equal in all coexisting phases. Eliminating $\phi=\phi(x)$ through Eq.~\eqref{eq:phi}, chemical potential and pressure are functions of composition alone with compact expressions
\begin{gather}
  \mu(x) = T\ln\frac{\phi(1-x)}{1-\phi} + f_1, \\ p(x) = -\frac{1}{2}(\phi x)^2 - T\ln(1-\phi).
\end{gather}
Through numerical root finding we determine the compositions $x^\pm$ obeying $\mu(x^+)=\mu(x^-)$ and $p(x^+)=p(x^-)$ simultaneously.

To determine the mean-field critical point, we inspect the derivative of the pressure
\begin{equation}
  \td{p}{x} = T\frac{T-x\phi+(x\phi)^2}{x^2(1-x)(1-\phi)},
\end{equation}
which has zeros for
\begin{equation}
  x\phi = \frac{1}{2}\pm\frac{1}{2}\sqrt{1-4T}.
\end{equation}
This implies the critical mean-field temperature $T_\text{c}^\text{MF}=\frac{1}{4}$ independent of $h$ (as expected for the Ising model). Solving $x_\text{c}\phi(x_\text{c})=\tfrac{1}{2}$ with Eq.~\eqref{eq:phi} yields
\begin{equation}
  \label{eq:crit}
  x_\text{c} = \frac{1}{1+e^{4h_\text{c}-2}}, \qquad
  \phi_\text{c} = \frac{1}{2x_\text{c}}.
\end{equation}
For the critical point to lie within the accessible range, we need $x_\text{c}\geqslant\tfrac{1}{2}$ and thus $h_\text{c}=h(T_\text{c})\leqslant\tfrac{1}{2}$.

\subsection{Monte Carlo simulations}

We also perform Monte Carlo simulations based on the lattice free energy Eq.~\eqref{eq:F:lg}. Picking a random molecule, we perform one of two trial moves: attempting to move the molecule to one of the $z=4$ neighboring lattice sites with probability $1-p_\text{conv}$ or switching to the other conformation with $p_\text{conv}$. Trial moves are accepted with probability
\begin{equation}
  \min\{1,e^{-\Delta\mathcal F/\kT}\}
\end{equation}
and rejected otherwise, where $\Delta\mathcal F=\mathcal F_\text{new}-\mathcal F_\text{old}$ is the change of the free energy Eq.~\eqref{eq:F:lg}. Attempted jumps to occupied sites are always rejected. The only parameter of the Monte Carlo algorithm is the probability $p_\text{conv}$, which does not influence averages in thermal equilibrium. Most simulation data has been obtained for $p_\text{conv}=0.1$.

To study direct phase coexistence, we employ an elongated simulation box with $K=L_x\times L_y$ lattice sites setting $L_y=40$ and $L_x=3L_y$. Due to the line tension, the average interface between dense and dilute domains aligns with the shorter box edge so that translational invariance holds along the $y$-axis but is broken along the $x$-axis. This facilitates the sampling of $x$-dependent averages, from which we extract the bulk properties away from the interfaces.

%% ---- discussion ----

\section{Discussion}

\subsection{Constant free energy difference}

We now discuss the phase diagrams following from the analysis of the previous section. To this end, we require an expression for $h(T)$, and thus for the constrained free energies of the conformations. For our purposes, we decompose
\begin{equation}
  f_\al = \eps_\al - Ts_\al,
\end{equation}
where $\eps_\al$ can be interpreted as binding energies with the substrate and the functions $s_\al$ describe the different entropies of the two conformations (in units of $k_\text{B}$).

\begin{figure}[t]
  \centering
  \includegraphics{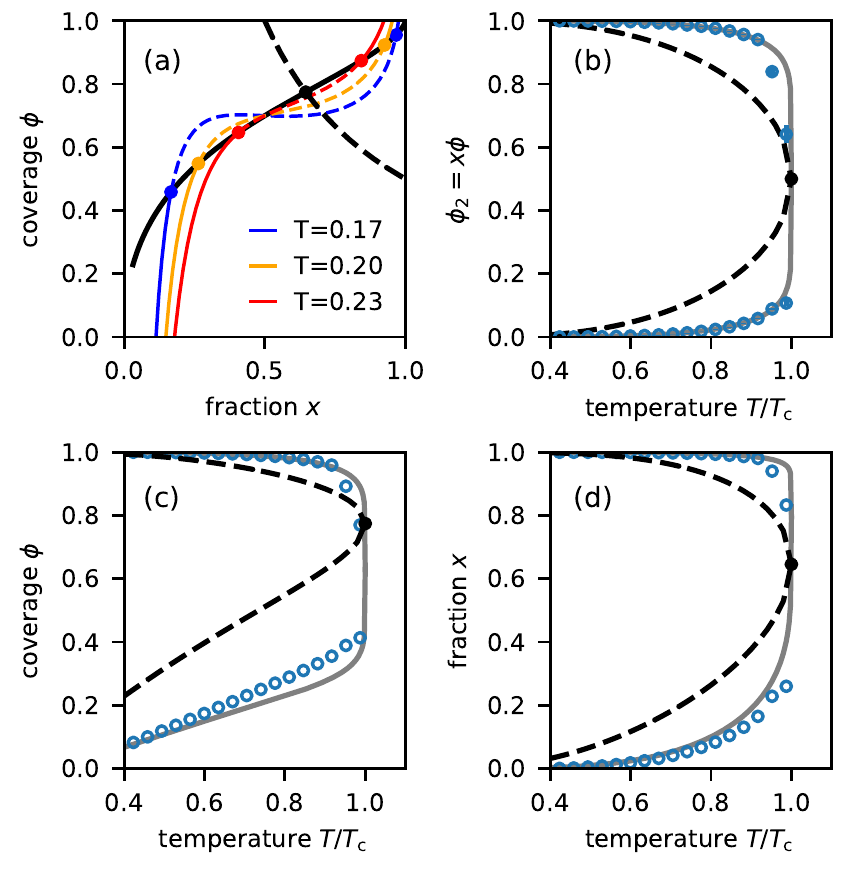}
  \caption{Phase diagram for constant $h=\eps=0.35$ ($s=0$). (a)~Mean-field equation of state $\phi(x;T)$ [Eq.~\eqref{eq:phi}] for three temperatures (colored lines). The critical point (black symbol) lies on the dashed black line [Eq.~\eqref{eq:crit}] and is determined by the value of $h_\text{c}=h$ [Eq.~\eqref{eq:crit}]. Colored symbols indicate the coexistence values, which form the binodal (solid black line). The unstable part of $\phi(x;T)$ is drawn as a dashed line. (b)~The coexisting values $\phi_2^\pm$ for the coverage $\phi_2=x\phi$ of standing molecules bounding the two-phase region. Dashed line is the mean-field result. Symbols indicate numerical results with the solid gray line showing the Ising prediction Eq.~\eqref{eq:ising}. The temperature is reduced by the corresponding critical temperature. (c,d)~Similar to (b) but showing (c)~the total coverage $\phi$ and (d)~the fraction $x$ of standing molecules.}
  \label{fig:pd:simple}
\end{figure}

The mean-field equation of state [Eq.~\eqref{eq:phi}] is plotted in Fig.~\ref{fig:pd:simple}(a) for constant $h=0.35$ (i.e., $s=0$) and several temperatures $T$ below the critical temperature $T_\text{c}$. Following an isotherm through increasing the global coverage $\phi$, the system is a disordered gas of lying molecules with a small fraction $x$ of standing molecules that is slowly increasing. At some $\phi^-$ we hit the binodal and now dense domains of standing molecules (large $\phi^+$ and $x^+$) start to coexist with dilute regions (small $\phi^-$ and $x^-$). Further increasing the global coverage $\phi$, these values remain the same but domains occupy a larger fraction $\kap$ of the system according to the lever rule
\begin{equation}
  \kap(\phi) = \frac{\phi - \phi^-}{\phi^+ - \phi^-}.
\end{equation}
Above $\phi\geqslant\phi^+$, the system is in the homogeneous dense (``liquid'') phase. Increasing the temperature $T$, the coexistence values approach each other (along the binodal) and the range of global occupations $\phi^-<\phi<\phi^+$ for which coexistence is possible shrinks and eventually vanishes at $T_\text{c}$.

In Fig.~\ref{fig:pd:simple}(b), we plot the coverage $\phi_2^\pm$ of standing molecules, which is symmetric. We also plot numerical coverages from Monte Carlo simulations, which agree qualitatively but approach each other more slowly at higher temperatures. The critical temperature in the simulations is that of the Ising model, $T_\text{c}^\text{MC}=[2z\ln(1+\sqrt 2)]^{-1}\simeq0.142$. We find that the numerically determined coverages follow Onsager's result for the coexisting densities~\cite{onsager44}
\begin{equation}
  \phi_2^\pm = \frac{1}{2} \pm \frac{1}{2}\left[1-\sinh\left(\frac{1}{2zT}\right)^{-4}\right]^{1/8}
  \label{eq:ising}
\end{equation}
(obtained in the thermodynamic limit). The standing molecules thus behave as predicted by the standard Ising model independent of $h$, which controls the total fraction of standing molecules.

We can improve upon our mean-field results through using that the chemical potential of standing molecules at coexistence reads $\mu_2^\ast=f_2-1/2$ in the thermodynamic limit (zero external field in the corresponding spin representation). For the free energy of the lying molecules, we use $F_1(N_1,K_1)=N_1f_1+F_0(N_1,K_1)=K_1[\xi f_1+Tf_0(\xi)]$ with fraction $\xi\equiv N_1/K_1$. At equilibrium, the chemical potentials
\begin{equation}
  \mu_1 = \left.\pd{F_1}{N_1}\right|_{K_1} = f_1+T\ln\frac{\xi}{1-\xi} \overset{!}{=} \mu_2^\ast
\end{equation}
have to be equal. This can be solved for the fraction $\xi=[1+e^{-(h-1/2)/T}]^{-1}$ as a function of temperature $T$ and free energy difference $h$ per molecule. Using the definition
\begin{equation}
  \xi = \frac{N_1}{K_1} = \frac{N-N_2}{K-N_2} = \frac{\phi-\phi_2}{1-\phi_2}
  \label{eq:xi}
\end{equation}
we finally obtain a prediction for the total coverage
\begin{equation}
  \phi^\pm = \xi + (1-\xi)\phi_2^\pm
  \label{eq:coex:ising}
\end{equation}
in both the dilute and dense phase. Employing Eq.~\eqref{eq:ising}, this prediction is shown in Fig.~\ref{fig:pd:simple}(c) together with the numerical results. While not perfect, the agreement improves dramatically over the simple mean-field ansatz Eq.~\eqref{eq:mf}. Finally, in Fig.~\ref{fig:pd:simple}(d) we show the fractions $x^\pm=\phi^\pm/\phi_2^\pm$ of standing molecules in both phases.

\subsection{Constant entropy difference}

\begin{figure}[t]
  \centering
  \includegraphics{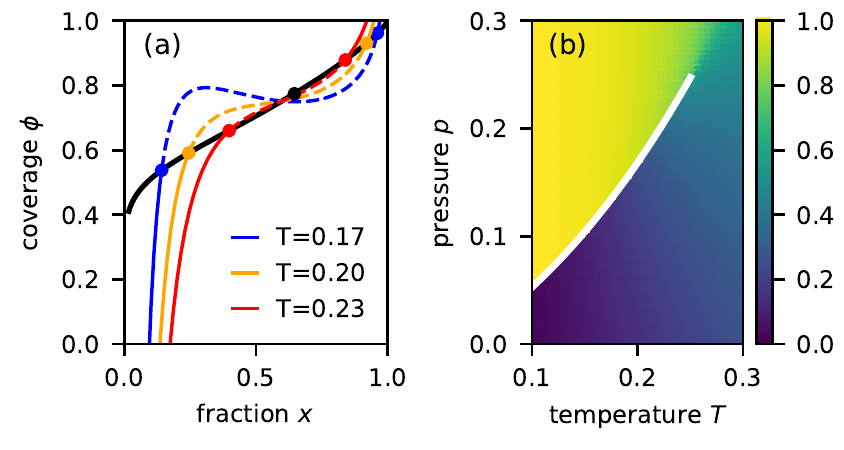}
  \caption{(a)~Mean-field phase diagram for temperature-dependent $h(T)=\eps-Ts$ with $\eps=0.45$ and $s=0.4$ (same critical point as in Fig.~\ref{fig:pd:simple}). (b)~Same parameters but plotted as a function of the intensive variables temperature $T$ and pressure $p$. The white line is the phase boundary ending in a critical point. The color indicates the fraction $x$ of standing molecules.}
  \label{fig:pd:normal}
\end{figure}

We now lift the degeneracy of the conformational entropies with temperature-dependent $h(T)=\eps-Ts$, where $\eps\equiv\eps_2-\eps_1$ is the difference of substrate binding energies and $s\equiv s_2-s_1$ the difference of internal entropy of the two molecular conformations. For $\eps>0$ conformation 2 binds more weakly to the substrate compared with conformation 1. A difference $s>0$ implies that the internal entropy of conformation 2 is larger, \emph{i.e.}, more internal modes (vibrations, etc.) are thermally excited.

Figure~\ref{fig:pd:normal}(a) shows the mean-field phase diagram for $\eps=0.45$ and constant $s=0.4$, implying the critical value $h_\text{c}=0.35$. While the binodal shifts up, qualitatively we observe a similar behavior to that shown in Fig.~\ref{fig:pd:simple} for constant $h$. In Fig.~\ref{fig:pd:normal}(b), we plot the same phase diagram but now for the intensive variables temperature and pressure. The coexistence region becomes a line, crossing which the coverage and fraction jump discontinuously.

\begin{figure}[b!]
  \centering
  \includegraphics{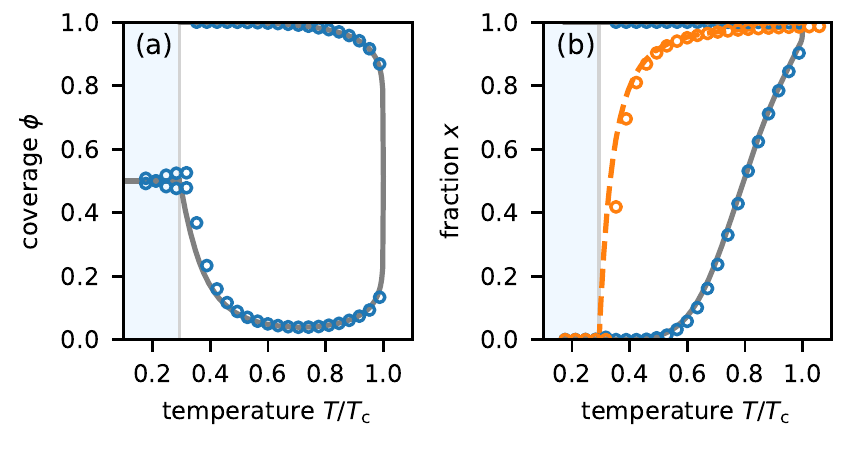}
  \caption{(a)~Phase diagram for $\eps=0.75$ and $s=6$ (these are the values employed for the simulation snapshots shown in Fig.~5 of Ref.~\citenum{aeschlimann21}) and global coverage $\phi=\tfrac{1}{2}$. Solid gray line is again the theoretical prediction and symbols indicate numerical results. (b)~Fraction of standing molecules (conformation 2, blue symbols and solid gray line) in the coexisting phases. The orange line and symbols show the total fraction $x$ of standing molecules. Below $T_0$ (vertical line), all molecules are found in the mobile lying conformation.}
  \label{fig:pd:angew}
\end{figure}

In the next step, we make the temperature dependence more pronounced through increasing both $\eps$ and $s$. Figure~\ref{fig:pd:angew}(a) now shows a qualitatively very different behavior with a non-monotonic $\phi^-$ that increases for lower temperatures. Cooling the substrate for sufficiently low global coverage thus leads to a reentrance into the homogeneous phase. We can easily rearrange Eq.~\eqref{eq:xi} to obtain the total fraction
\begin{equation}
  x = \frac{1-\xi/\phi}{1-\xi}
  \label{eq:x}
\end{equation}
of standing molecules (conformation 2). In Fig.~\ref{fig:pd:angew}(b) we show that this fraction declines and reaches zero at the non-zero temperature $T_0$. The system is always homogeneous below $T_0$ with all molecules being mobile (conformation 1). Specifically, for $\phi=\tfrac{1}{2}$ we find $T_0=(\eps-\tfrac{1}{2})/s$. For $\eps<\tfrac{1}{2}$ we have conventional coexistence without reentrance ($T_0<0$). For $s>0$ and $\tfrac{1}{2}\leqslant\eps\leqslant\tfrac{1}{2}+T_\text{c}s$ we observe reentrance, whereby at lower temperatures the energetic gain of lying molecules overcomes the entropy surplus of standing molecules. For even larger $\eps$, the system is always homogeneous with a transition from standing to lying molecules at $T_0>T_\text{c}$.

\subsection{Temperature-dependent entropy}

\begin{figure}[b!]
  \centering
  \includegraphics{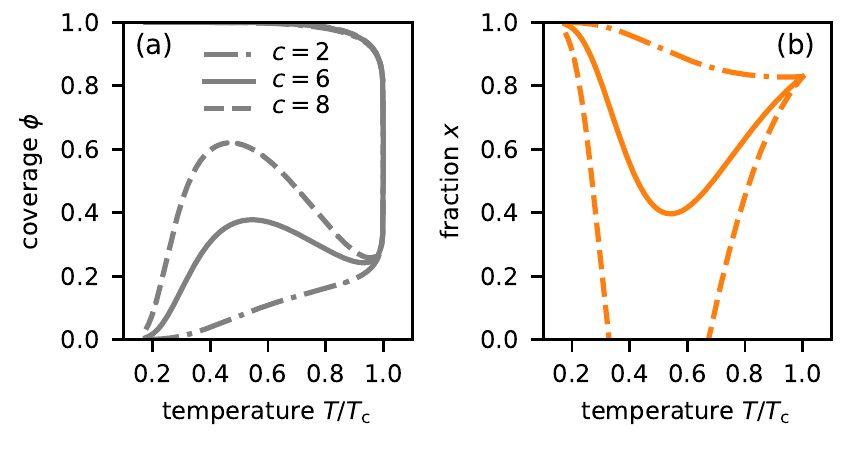}
  \caption{(a)~Coexisting coverages $\phi^\pm$ from Eq.~\eqref{eq:coex:ising} using Eq.~\eqref{eq:h:heatcap} for $h(T)$ at global coverage $\phi=\tfrac{1}{2}$, $\eps=0.25$, and $s=0$ for three molecular heat capacities $c$. (b)~Corresponding fraction $x$ of standing molecules. For $c=8$ the standing molecules again vanish, entering a homogeneous state of mobile lying molecules, but reappear at lower temperatures.}
  \label{fig:pd:heatcap}
\end{figure}

Assuming temperature-independent internal entropies $s_\al$ might not be appropriate for complex molecules. Close to the critical temperature $T_\text{c}$, we expand $s_\al(T)=s_\al(T_\text{c})+(c_\al/T_\text{c})(T-T_\text{c})$ with heat capacities $c_\al=c_\al(T_\text{c})$ per molecule in conformation $\al$ (again measured in units of $k_\text{B}$). The free energy difference per molecule now reads
\begin{equation}
  h(T) = f_2 - f_1 = \eps - Ts - T(T/T_\text{c}-1)c
  \label{eq:h:heatcap}
\end{equation}
with $s$ the entropy difference at $T_\text{c}$ and $c\equiv c_2-c_1$ the difference in molecular heat capacities. Figure~\ref{fig:pd:heatcap}(a) demonstrates that the resulting phase diagram again exhibits reentrance into the homogeneous phase for sufficiently large $c$. The total fraction of molecules in conformation 2 drops as before but then rebounds at lower temperatures do to the decrease of the entropy difference [Fig.~\ref{fig:pd:heatcap}(b)]. Increasing $c$ even further leads to the vanishing of standing molecules and an intermediate homogeneous phase before again ordering at low temperatures. The temperature range $T_0^\pm$ of this intermediate phase can be calculated from the vanishing numerator of Eq.~\eqref{eq:x}. Specifically, for $s=0$ and $\phi=\tfrac{1}{2}$ we find the simple expression
\begin{equation}
  T_0^\pm = \frac{1}{2}T_\text{c}\left[1\pm\sqrt{1+4\frac{\eps-\tfrac{1}{2}}{cT_\text{c}}}\right]
\end{equation}
with critical value $c_\ast\simeq7.05$ above which the system goes through the intermediate homogeneous phase.

\subsection{Condensation or freezing?}

Before concluding we briefly comment on a subtle point. Both the freezing into a solid and the condensation into a denser liquid phase can be described as order-disorder transitions. In contrast to condensation, freezing in three dimensions involves the breaking of translational symmetry into a discrete crystal symmetry. Two-dimensional systems can still freeze into a solid but without truly long-range \emph{positional order}, which is destroyed by thermal fluctuations~\cite{mermin66}. Strictly speaking, therefore, no crystal exists in two dimensions. While a molecular monolayer comes very close to two dimensions, the underlying substrate induces an external potential that, even if weak, breaks translational symmetry and thus can reintroduce long-range order~\cite{strandburg88}. Note that a lattice model is insufficient to capture the spontaneous breaking of translational symmetry since the discrete lattice cannot accommodate a continuous symmetry. Rather than freezing into a periodic crystal, our model thus describes condensation of molecules into dense domains.

%% ---- conclusions ----

\section{Conclusions}

The Ising model provides a rigorous statistical mechanics underpinning for a wide class of disorder-order transitions and serves as a paradigm for reversible structure formation. It captures the competition between short-range isotropic attraction and translational entropy, describing the condensation of a dense liquid coexisting with a dilute gas. Crossing a first-order transition gives rise to phase ordering kinetics that can be exploited to control and create non-equilibrium structures and morphologies~\cite{dong21}.

In contrast to atoms and colloidal particles, molecules have internal degrees of freedom. For molecules with approximately isotropic intermolecular interactions, the Ising model is still an appropriate large-scale description but the ``field'' may acquire a dependence on temperature due to the intramolecular entropy. An additional competition between intramolecular energy and entropy opens a route to reshape the phase diagram. Here we have studied the case of two molecular conformations, whereby intermolecular interactions are only assumed for one confirmation (here conformation 2). This restriction has allowed us to exploit the analytical solution of the Ising model in two dimensions to accurately construct the phase behavior, which is determined by two temperatures: the critical temperature $T_\text{c}$ of the Ising model below which domains form (driven by the interactions of conformation 2) and the temperature $T_0$ below which conformation 2 vanishes. While $T_\text{c}$ is fixed, tuning $T_0$ allows to shape the coexistence region and to find parameters for which reentrance into a homogeneous disordered phase is possible.

Such reentrance upon cooling has been observed recently for dimolybdenum tetraacetate and has been attributed to two different molecular conformations~\cite{aeschlimann21}: Lying molecules (conformation 1) bind more strongly due to the metal--metal interaction ($\eps_1<\eps_2$), which at the same time suppresses internal degrees of freedom ($s_1<s_2$) due to steric hindrance with the substrate (cf. Fig.~\ref{fig:conf}). To relate the scenario of inverse condensation to the experiments of Ref.~\citenum{aeschlimann21}, we first note that (for the ordered phase to be present) room temperature needs to be below the the critical temperature. This condition implies the lower bound $\eb>25.7\unit{meV}/(zT_\text{c})\approx45\unit{meV}$. As reference, the effective binding energy of C$_{60}$ molecules has been estimated to be $\eb\approx235\unit{meV}$~\cite{janke20} and thus is five times larger. The difference in substrate binding energies between the two conformations of dimolybdenum tetraacetate has been determined to $\tilde\eps\approx410\unit{meV}$~\cite{kollamana18}, hence dimensionless $\eps=\tilde\eps/(z\eb)$ of order unity seem reasonable. As an alternative scenario, we have considered a difference of heat capacities between both conformations, which also leads to reentrant behavior for differences exceeding a few $k_\text{B}$'s. Our results thus indicate that inverse condensation should indeed be observable for a wide range of adsorbed organic molecules.

Beyond the phase behavior of adsorbed molecules, our results are potentially interesting for condensates of proteins forming membraneless organelles inside the cell~\cite{hyman14,boeynaems18}, which are studied intensively at the moment. Reentrant behavior has already been reported~\cite{banerjee17a,krainer21} and could be studied theoretically in on-lattice ``stickers-and-spacers'' models for multivalent proteins~\cite{michels22}.

%% ---- acknowledgments ----

\begin{acknowledgments}
  We thank Angelika Kühnle for fruitful discussions. We thank her and Joshua Robinson for a critical reading of the manuscript. We gratefully acknowledge financial support through the Deutsche Forschungsgemeinschaft: research grant 319880407 and Research Training Group GRK 2516 ``Control of structure formation in soft matter at and through interfaces'' (grant number 405552959).
\end{acknowledgments}

%% ---- author declarations ----

\section*{Author declarations}

\subsection*{Conflict of interest}

The authors have no conflicts to disclose.

\section*{Data availability}

The numerical data that support the findings of this study are available from the corresponding author upon reasonable request.

%% ---- bibliography ----

\end{document}